\documentclass[twocolumn,aps]{revtex4}
\usepackage{epsfig,pstricks,graphicx}
\usepackage{amsmath,amssymb}

\def\CC{{\rm\kern.24em \vrule width.04em height1.46ex depth-.07ex
\kern-.30em C}}
\def\RR{{\rm
         \vrule width.04em height1.58ex depth-.0ex
         \kern-.04em R}}
\def\id{{\rm 1\kern-.22em l}}
\def\e{{\rm e}}

\newcommand{\beq}{\begin{equation}}
\newcommand{\beqa}{\begin{eqnarray}}
\newcommand{\nbeqa}{\begin{eqnarray*}}
\newcommand{\eeq}{\end{equation}}
\newcommand{\eeqa}{\end{eqnarray}}
\newcommand{\neeqa}{\end{eqnarray*}}
\newcommand{\bra}[1]{\left\langle #1 \right |}
\newcommand{\ket}[1]{\left | #1 \right\rangle}

\begin{document}
\tighten

\title{Tangles of superpositions and the convex-roof extension}
\author{Andreas Osterloh$^1$}
\author{Jens Siewert$^{2}$}
\author{Armin Uhlmann$^{3}$}
\affiliation{
    $^1$ Institut f\"ur Theoretische Physik, 
         Leibniz Universit\"at Hannover, D-30167 Hannover, Germany\\
    $^2$ Institut f\"ur Theoretische Physik, 
         Universit\"at Regensburg, D-93040 Regensburg, Germany\\
     $^3$ Institut f\"ur Theoretische Physik, 
         Universit\"at Leipzig, D-04109 Leipzig, Germany
            }

\begin{abstract}
We discuss aspects of the convex-roof extension of multipartite 
entanglement measures, that is, $SL(2,\CC)$ invariant tangles.
We highlight two key concepts that contain valuable information about
the tangle of a density matrix: the {\em zero-polytope} is a convex set of 
density matrices with vanishing tangle whereas the
{\em convex characteristic curve} readily provides a non-trivial
lower bound for the convex roof and serves as 
a tool for constructing the convex roof outside the zero-polytope.
Both concepts are derived from the tangle for superpositions  
of the eigenstates of the density matrix.
We illustrate their application by considering examples of 
density matrices for two-qubit and three-qubit states of rank 2, 
thereby pointing out both the power and the limitations of the concepts.
\end{abstract}

\maketitle

\section{Introduction}

One of the challenges in present quantum information theory is the
quantification of multipartite entanglement both in pure and
mixed states~\cite{Plenio07}. Once a measure for pure
multipartite states is given (a ``tangle'' -- that is, a polynomial 
invariant, see below), the corresponding measure for 
mixed states can be obtained via the convex-roof 
extension~\cite{Uhlmann98}. Although its definition is straightforward,
the practical evaluation of the convex-roof extension is a difficult
mathematical problem. To date a general analytical method is known only for
the concurrence of two-qubit mixed states~\cite{Wootters98,Uhlmann00}.

The convex roof is obtained by minimizing the average tangle of a given
mixed state over all possible decompositions of that state into pure states. 
As all elements
in a decomposition are superpositions of the eigenstates of the
given density matrix, the minimal tangle will depend on the possible tangle
values for those superpositions.

Recently, several authors have discussed how the entanglement of 
superpositions of some given pure states is related to the entanglement
contained in those states~\cite{Linden05,Cavalcanti07,Gour07,Cerf07,Song07}.
Here, we are considering a different problem, namely how the entanglement
of a mixed state depends on the entanglement of the superposition of its
constituents. To this end, we introduce two 
concepts that
facilitate the evaluation of mixed-state tangles for rank-2 states: 
the {\em zero polytope}
and the {\em convex characteristic curve}. For a given mixed state $\rho$,
the zero polytope is the convex set of those density matrices which
have eigenstates in the range of $\rho$ and, at the same time,
have vanishing tangle.
On the other hand, the convex
characteristic curve is the largest convex
function (of a suitable parameter) that does not exceed the tangle 
of any pure state 
in the range of $\rho$ (for detailed explanation see below).

The outline of this article is as follows. First, we introduce some
terminology (Section II) and then give precise definitions of the zero 
polytope and the characteristic curve (Section III). There,
we also discuss implications
for the convex roof of the tangle in rank-2 states as well as the possible
extension of these concepts to ranks larger than two. In Section IV
we illustrate the application to examples of rank-2 states of two and 
three qubits.

\section{Basic concepts}
\subsection{Pure-state entanglement measures}
\label{pure-state-measures}

A function quantifying entanglement is required to be an
entanglement monotone~\cite{MONOTONES}. Its crucial property is
that it be non-increasing (on average) under stochastic local operations and
classical communication (SLOCC).
Stochastic local equivalence of two $N$-qubit states is represented
by convertibility under $SL(2,\CC)^{\otimes N}$ 
transformations~\cite{SLOCC,Duer00}, and a homogeneous function 
on $N$ qubits that is invariant under such operations is an entanglement 
monotone~\cite{VerstraeteDM03}.

Here we investigate entanglement measures related to the
concurrence introduced by Wootters and 
co-workers~\cite{Hill97,Wootters98,Coffman00}.

The concurrence $C(\phi_{AB})$ measures the degree of bipartite entanglement 
shared between the parties $A$ and $B$ in a pure two-qubit state 
$\ket{\phi_{AB}}\in{\cal H}_A\otimes{\cal H}_B$. In terms of the coefficients
$\{\phi_{00}, \phi_{01}, \phi_{10}, \phi_{11}\}$ of $\ket{\phi_{AB}}$ 
with respect
to an orthonormal basis of product states it is defined as
\begin{equation}
 C(\phi_{AB})\ =\ 2 | \phi_{00}\phi_{11}-  \phi_{01}\phi_{10} |
  \ \ .
\label{def_concurrence}
\end{equation}
The concurrence is maximal for Bell states such as 
\mbox{$(\ket{00}+\ket{11})/\sqrt{2}$} while it vanishes for factorized states.

A measure for three-party entanglement has been introduced in 
Ref.~\cite{Coffman00}. The 3-tangle $\tau_3(\psi)$ of a three-qubit
state $\psi\in {\cal H}_A\otimes{\cal H}_B\otimes{\cal H}_C$
can be expressed by using the wavefunction coefficients 
$\{\psi_{000},\psi_{001},\ldots,\psi_{111}\}$ as
\nbeqa
\label{def-3tangle}
\tau_3 &=& 4\ |d_1 - 2d_2 + 4d_3|\\
  d_1&=& \psi^2_{000}\psi^2_{111} + \psi^2_{001}\psi^2_{110} + \psi^2_{010}\psi^2_{101}+ \psi^2_{100}\psi^2_{011} \\
  d_2&=& \psi_{000}\psi_{111}\psi_{011}\psi_{100} + \psi_{000}\psi_{111}\psi_{101}\psi_{010}\\ 
    &&+ \psi_{000}\psi_{111}\psi_{110}\psi_{001} + \psi_{011}\psi_{100}\psi_{101}\psi_{010}\\
    &&+ \psi_{011}\psi_{100}\psi_{110}\psi_{001} + \psi_{101}\psi_{010}\psi_{110}\psi_{001}\\
  d_3&=& \psi_{000}\psi_{110}\psi_{101}\psi_{011} + \psi_{111}\psi_{001}\psi_{010}\psi_{100}\ \ .
\neeqa
For the GHZ state 
\begin{equation}
       \ket{GHZ}\ = \ \frac{1}{\sqrt{2}}\ (\ket{000}\ +\ \ket{111})
\label{ghz}
\end{equation}
the 3-tangle becomes maximal: $\tau_3(GHZ)=1$, and it 
vanishes for any factorized state. It is most remarkable that there is
a class of 
entangled three-qubit states for which $\tau_3$ vanishes, i.e.,
these states are not SLOCC equivalent to any GHZ-type state~\cite{Duer00}.
This class is represented by the $\ket{W}$ state
\begin{equation}
  \ket{W}\ =\ \frac{1}{\sqrt{3}}\ ( \ket{100}\ +\ \ket{010}\ +\ \ket{001} )
  \ \ ,
\label{wstate}
\end{equation}
and it is easy to check that $\tau_3(W)=0$.

Both concurrence and the 3-tangle are the unique two- and three-qubit 
versions of more general polynomial measures invariant under
local $SL(2,\CC)$ transformations~\cite{VerstraeteDM03,Leifer04}. 
These quantities are homogeneous
of even degree in the coefficients of the wavefunction (with respect
to a product basis) and
play an important role as fundamental $SL(2,\CC)$ invariants in 
invariant theory~\cite{BriandLT03}. Therefore they serve as 
class-selective measures of (multipartite) entanglement~\cite{OS04,OS05}.
In what follows, we refer to such polynomial invariants as {\em tangles}.

\subsection{Entanglement measures for mixed states -- convex-roof extension}
\label{mixed-state-measures}

Given a continuous real function $g$ on the space of pure states
$\Omega^{\mathbf{ex}}$, this function
can be extended to the space of mixed states in the 
following way~\cite{BennettDiVincenzo96,Benatti96,Uhlmann00}. 
Let $\Omega$ be the convex (and compact) 
set of normalized density operators. A state $\rho\in \Omega$
can be written as a convex combination
\begin{equation}
   \rho\ =\ \sum_j p_j \pi_j,\ \ \ \ \ \pi_j:=\frac{\ket{j}\!\bra{j}}
                                            {\langle j | j \rangle}
\end{equation}
where the pure states $\pi\in \Omega^{\mathrm{ex}}$ 
are extremal points of $\Omega$.
The real function $g^{\cup}: \Omega \to \mathbb{R}$ is a 
{\em convex-roof 
extension} of $g: \Omega^{\mathrm{ex}}\to \mathbb{R}$
if $g^{\cup}$ coincides with $g$
on $\Omega^{\mathrm{ex}}$ and
\beq
              g^{\cup}(\rho)\ :=\ \min \sum p_j g(\pi_j) \ \ .
\label{defroof}
\eeq
where the minimum is taken over all decomposition of $\rho$.
With this in mind it is formally straightforward to extend the 
definition of the concurrence to mixed two-qubit states $\rho$ with
a decomposition $\rho = \sum p_j \pi_j^{AB}$ into pure
two-qubit states $\pi_j^{AB}$ of the parties $A$ and $B$
\beq
           C(\rho)\ =\ \min \sum p_j C(\pi_j^{AB})\ \ .
\label{mixed_concurrence}
\eeq
Correspondingly, we have the 3-tangle 
for mixed three-qubit states $\rho=\sum p_j \pi_j^{ABC}$
\beq
           \tau_3(\rho)\ =\ \min \sum p_j \tau_3(\pi_j^{ABC})
\label{mixed_3tangle}
\eeq
where $\pi_j^{ABC}$ denote pure three-qubit states. 
A decomposition for which the minimum of the respective function
is realized, is called an {\em optimal decomposition}.

\section{Entanglement of superpositions and implications for the convex roof}

It is in general  difficult to evaluate the convex-roof extension, i.e.\
the tangle of a given density matrix $\rho$. 
To this end, it is required
to study the entanglement of pure states in the range of $\rho$, 
${\cal R}_\rho$, as can be seen from Definition (\ref{defroof}).
Our starting point is the observation  that the pure states 
$\ket{\psi}_\rho \in {\cal R}_\rho$
are superpositions of the eigenstates of $\rho$.

\subsection{Density matrices with vanishing tangle: the zero-polytope}

An important question is whether or not a tangle $\tau$ vanishes for a
given  $\rho$. That is, we ask whether there is
a decomposition of $\rho$ into pure 
states $\{\ket{\psi_j}_\rho\}$ 
such that $\tau(\ket{\psi_j}_\rho)=0$ for all $j=0,\ldots,r^2-1$ where 
$r$ is the rank of $\rho$.
A polynomial $SL(2,\CC)$ invariant 
$\tau$ is a homogeneous function of the coefficients of $\ket{\psi}_\rho$.
The homogeneous degree 
of $\tau$ (denoted by $h$) is an even positive integer. 
Writing $\ket{\psi}_\rho =\sum_{j=0}^{r-1}z_j \ket{\psi_j}$, 
this implies that $\tau(\ket{\psi}_\rho)$ is a polynomial of degree
$h$ in the complex variables $z_0, \dots , z_{r-1}$.
The situation is most easily understood by considering
rank-2 density matrices.
Let us assume there is at least one eigenstate
$\ket{\psi_1}$ of $\rho$ such that $\tau(\psi_1)\neq 0$
(otherwise we have trivially $\tau(\rho)=0$).
Since normalization does not play any role for this question,
we then write
\beq\label{psi-rho-z}
\ket{\psi}_\rho = \ket{\psi_0}+ z \ket{\psi_1}\ \ .
\eeq
The tangle is then a polynomial of degree $h$ of a single complex 
variable $z$ and consequently has precisely $h$ zeros
$\zeta_{1},\dots,\zeta_h$.
If both states have zero tangle we have to add the solution $z=\infty$ 
to the $h-1$ finite solutions of $\tau(\ket{\psi}_\rho)=0$.
These zeros\footnote{For real states $\ket{\psi}_\rho$ the 
zeros are either real or appear in complex conjugate pairs.}
correspond to $h$ pure states
\nbeqa
\ket{Z_1}_\rho&=& \ket{\psi_0}+\zeta_1\ket{\psi_1}\\
&\vdots&\\
\ket{Z_h}_\rho&=& \ket{\psi_0}+\zeta_h\ket{\psi_1}
\neeqa
satisfying $\tau(Z_j)=0$ for all $j=1,\dots,h$.
They span a polytope with $h$ corners that contains precisely those
density matrices which can be decomposed into  these states, i.e.,
all convex combinations of 
$\pi_{{}_{Z_1}}, \dots,\pi_{{}_{Z_h}}$.
All the density matrices in this polytope have zero tangle (and outside
there is no other state with zero tangle) and therefore
we  call it the {\em zero-polytope};
its occurrence is generic for all polynomial $SL(2,\CC)$-invariant
entanglement monotones. This includes the concurrence, 3-tangle
and multipartite entanglement monotones constructed in~\cite{OS04,OS05}. 

We will now discuss briefly how the concept of the zero-polytope can be
extended to higher-rank density matrices.
As the tangle of homogeneous degree $h$ evaluated on the pure 
state gives a polynomial of degree $h$
in the variables $z_j$,
for $z_2,\dots, z_{r-1}$ kept fixed, we return to
the situation explained above, and the remaining polynomial has precisely
$h$ zeros, $\zeta_j(z_2,\dots, z_{r-1})$ with corresponding states
\[
\ket{Z_j;z_2,\dots, z_{r-1}} =
          \ket{\psi_0}+\zeta_j(z_2,\dots, z_{r-1})\ket{\psi_1}
           +\sum_{k=2}^{r-1} z_j\ket{\psi_j}\  .
\]
Upon varying $z_2,\dots, z_{r-1}$, each of these states will describe
an $2(r-2)$ dimensional 
(real) manifold in the $r^2-2$ dimensional 
manifold of states $\ket{\psi}\bra{\psi}$. 
The convex hull of the union of these $h$ manifolds
contains exactly all density matrices in ${\cal R}_\rho$
with zero tangle.\\
Below we will discuss the zero-polytope for
two explicit examples with rank $r=2$.

\subsection{The convex characteristic curve}

It is well-known~\cite{Schroedinger1936} that from the decomposition 
of a rank-$r$ density matrix
$\rho=\sum_{j=0}^{r-1} p_j \ket{\psi_j}\!\bra{\psi_j}$
into its eigenstates $\{\ket{\psi_j}\}$,
any other decomposition $\rho=\sum_l \ket{\chi_l}\!\bra{\chi_l}$
of length $m \geq r$ can be obtained with a unitary 
\mbox{$m \times m$} matrix $U_{lj}$ 
via $\ket{\chi_l}\ = \ \sum_{j=0}^{m-1} U_{lj} \sqrt{p_j}\ket{\psi_j}$.
Therefore, the tangle of all pure states $\ket{\psi}_\rho$
\[
     \ket{\psi}_\rho \ =\ \sum_{j=0}^{r-1} c_j \ket{\psi_j}\ \ ,\ \ \ \ \  \sum_j|c_j|^2=1
\]
can be regarded as a finger print of the tangle for {\em all} density
matrices in ${\cal R}_\rho$.
This suggests to look for the minimum tangle of states $\ket{\psi}_\rho$;
but rather than taking the global minimum, which is zero,
we will look for local minima when the weights in the superposition
are kept fixed. 
The degrees of freedom for this minimization
are the relative phases in the superposition. 
In an appropriate geometric representation, these 
minima describe characteristic curves of the tangle as a function
of the weights.

Let us discuss the simplest case first, that is, a rank-2 density matrix
with its eigendecomposition
\beq
\rho = p_0 \ket{\psi_0}\!\bra{\psi_0} +  p_1 \ket{\psi_1}\!\bra{\psi_1}
\eeq

with $p_j\in\RR$, $0\le p_j\le 1$ such that $p_0+p_1 =1$.

All pure states in ${\cal R}_\rho$ can be written as
\beq
\ket{\psi}_\rho = \sqrt{q}\ket{\psi_0}+ \sqrt{1-q}\e^{i\varphi}\ket{\psi_1}
\label{superpos}
\eeq
where $q, \varphi\in\RR$ and $0\le q\le 1$, $0\le \varphi\le 2\pi$.
We may consider the tangle $\tau$ of such a state 
(for a given phase $\varphi$) as a function of $q$ in a $\tau-q$ diagram 
(cf.\ also Fig.~1).
This function represents a ``characteristic curve'' for $\tau$ 
and is in general not convex.

Clearly, each point $(q,\tau(\psi(q,\varphi)))$ in the $\tau-q$ plane stands 
for the tangle of the pure state projector 
$\ket{\psi(q,\varphi)}_\rho\!\bra{\psi(q,\varphi)}$. Now consider
an alternative length-2 decomposition of $\rho$,  
$\rho = \tilde{p}_0  \ket{\phi_0}_\rho\!\bra{\phi_0}
                             + \tilde{p}_1  \ket{\phi_1}_\rho\!\bra{\phi_1}$.

The states $\ket{\phi_0}_\rho$ and $\ket{\phi_1}_\rho$ have the we
weights $q_0$, $q_1$ and the relative phases $\varphi_0$ and $\varphi_1$.
As these states form a decomposition of $\rho$ their parameters obey
$p_0=\tilde{p}_0 q_0 +\tilde{p}_1 q_1$. Moreover, the phases need to
be adjusted properly.
Note that, for arbitrary phases, a density matrix 
$p_0  \ket{\psi_0}_\rho\!\bra{\psi_0}
 + p_1  \ket{\psi_1}_\rho\!\bra{\psi_1} 
 + \alpha\ket{\psi_0}_\rho\!\bra{\psi_1} + {\rm h.c.}$ is obtained which,
in the eigenbasis, has the correct diagonal elements.

The average tangle of the decomposition 
$\rho = \tilde{p}_0  \ket{\phi_0}_\rho\!\bra{\phi_0}
                             + \tilde{p}_1  \ket{\phi_1}_\rho\!\bra{\phi_1}$
is visualized 
in the $\tau-q$ plane by connecting the points 
$(q_0,\tau(q_0,\varphi_0))$ and $(q_1,\tau(q_1,\varphi_1))$  
corresponding to the tangle of the states $\ket{\phi_0}_\rho$,
$\ket{\phi_1}_\rho $ by a straight line.

Its value $\tilde{p}_0 \tau(\phi_0) +\tilde{p}_1 \tau(\phi_1)$
is assumed for the abscissa value 
$q\equiv p_0=\tilde{p}_0 q_0 + \tilde{p}_1 q_1$.
The generalization to decompositions of
$\rho$ of length $>2$ is straightforward.
Hence, a given decomposition of $\rho$ 
can be assigned a point in the $\tau-q$ plane 
with abscissa $p_0$ and ordinate equal to the average tangle $\tau$ 
that results from convexly combining the tangles of the

pure states in the decomposition.

\begin{figure}[h]
\includegraphics[width=.45\textwidth]{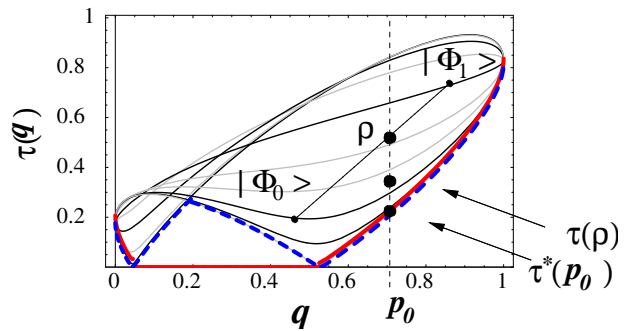}
\caption{Characteristic curves $\tau(q,\varphi)$ 
          in the $\tau - q$ plane (solid gray lines)
          for the superpositions in Eq.~(\ref{superpos}).
          The (blue) dashed line represents the minimum $\tilde{\tau}(q)$
          from Eq.~(\ref{tautilde})
          and the solid (red) line its function convex hull
          $\tau^{\ast}(q)$,
          i.e., the convex characteristic curve (for details see text).\newline
          The point with label $\rho$ indicates the average tangle of
          a particular length-2 decomposition of the density
          matrix $\rho=\tilde{p}_0\ket{\phi_0}
          +\tilde{p}_1 \ket{\phi_1}$. Moreover we illustrate the actual tangle
          of this density matrix $\tau(\rho)$ and its lower bound
          $\tau^{\ast}(p_0)$ according to the convex characteristic curve.}
          
\label{kennlinien1}
\end{figure}

The actual value of the tangle $\tau(\rho)$ is the minimum for all 
possible decompositions.
In our geometric visualization this means to find the smallest 
possible $\tau(p_0)$ that may
result from a convex combination of at most $r^2$ pure states. 
In order to find this value of $\tau$ 
(or at least approximate it) one
may try the following approach: first minimize the pure-state 
tangle $\tau(q,\varphi)$ for each argument $q$ 
\beq
   \tilde{\tau}(q)\ = \ \min_{\varphi}{[\tau(\psi(q,\varphi))]}
\label{tautilde}
\eeq
and then try to find 
decompositions from these minimal values $\tilde{\tau}(q)$.
That is, we need to investigate convex combinations of pure states and their corresponding
tangle. However, we note that the function $\tilde{\tau}(q)$ is in general not convex. 

Therefore we define
$\tau^{\ast}(q)$ as the function convex hull of $\tilde{\tau}(q)$ 
(i.e., the largest convex function that does not exceed $\tau(q,\varphi)$)
and call it the {\em convex characteristic curve}.
It is not difficult to see that the convex characteristic curve 
provides a non-trivial lower bound for the 
tangle $\tau(\rho)$ of the initial rank-2 density matrix $\rho$. 
Any (normalized) vector $\ket{\phi_j^{\mathrm{opt}}}_\rho$ in an optimal
decomposition of 
$\rho=\sum p_j^{\mathrm{opt}} \ket{\phi_j^{\mathrm{opt}}}_\rho\!\bra{\phi_j^{\mathrm{opt}}}$ 
is
characterized by some $q_j$ and a phase $\varphi_j$ in analogy with Eq.~(\ref{superpos}). 
A lower bound for the tangle of the pure state 
$\ket{\phi_j^{\mathrm{opt}}}_\rho\!\bra{\phi_j^{\mathrm{opt}}}$ is given by the 
convex characteristic curve:
$\tau(\phi_j^{\mathrm{opt}})\ge \tau^{\ast}(q_j)$. 
Consequently, the tangle of $\rho$ is
bounded from below by the convex combination 
$\sum p^{\mathrm{opt}}_j \tau^{\ast}(q_j)$
of the points $\{(q_j,\tau^{\ast}(q_j))\}$ on the convex characteristic
curve. As the weight of $\ket{\psi_0}$ in $\rho$ equals $p_0$ we find
\beq
        \tau(\rho)= \sum p_j^{\mathrm{opt}} \tau(\phi_j^{\mathrm{opt}})\ge \sum p_j^{\mathrm{opt}} \tau^{\ast}(q_j) \geq \tau^{\ast}(p_0) \ . 
\label{abschaetztau}
\eeq
This is a tight estimate of $\tau(\rho)$. 
The equal sign in the first inequality in Eq.~(\ref{abschaetztau}) 
holds if and only if there is an optimal decomposition such that all points 
corresponding to the states of this decomposition lie on the 
convex characteristic curve.
For the equal sign in the second inequality in Eq.~(\ref{abschaetztau}), 
either $\tau^*$ must be affine or all $q_i$ have to coincide.

If the rank of the given density matrix is larger than two, the convex characteristic
curve can be generalized to a convex manifold. 
Suppose we have a family of rank-$r$
density matrices with its eigendecomposition
\beq
    \rho
            \ =\ p_0 \ket{\psi_0}\!\bra{\psi_0}+\ldots+p_{r-1}\ket{\psi_{r-1}}\!\bra{\psi_{r-1}}
\label{rankrrho}
\eeq
and $\sum_{j=0}^{r-1}p_j=1$.
Then we define
\begin{widetext}
\beqa
\label{char-curve-gen}
\tilde{\tau}(q_0,\dots,q_{r-2})&=&\min_{\vec{\varphi}}\tau[\psi(q_0,\dots,q_{r-2};\varphi_1,\dots,\varphi_{r-1})]\\
\ket{\psi(q_0,\dots,q_{r-2};\varphi_1,\dots,\varphi_{r-1})}&=&
                    \sqrt{q_0}  \;                   \ket{\psi_0}
  +\sum_{j=1}^{r-1} \sqrt{q_j}\; \e^{i \varphi_{j}} \ket{\psi_j}
\eeqa
\end{widetext}
with $q_j\in\RR$ such that $\sum_{j=0}^{r-1} q_j =1$. The convex characteristic
manifold $\tau^{\ast}(q_0,\ldots,q_{r-2})$ is then defined as
the function convex hull of $\tilde{\tau}(q_0,\ldots,q_{r-2})$.

In analogy with the rank-2 case, the value of $\tau^{\ast}(p_0,\ldots,p_{r-2})$
gives a strict lower bound for the tangle of the density matrix in 
Eq.~(\ref{rankrrho}).
However, due to its higher dimensionality
it is much more difficult than in the rank-2 case to completely 
characterize this manifold and use it for estimates. 
Even as a tool for a numerical or graphical method it will be
rather hard to handle.

In order to illustrate the relevance of the convex characteristic curves/manifolds, 
but to also highlight caveats, we discuss two examples on 
specific rank-2 density matrices for two and three qubits.

\section{Examples}

In the following we illustrate the concepts introduced above in
two examples: three-qubit mixtures of GHZ and W states~\cite{LOSU}, and a
generic two-qubit example. 
While in the highly symmetric three-qubit case the
convex characteristic curve leads directly to an
analytic solution for the convex roof of a
whole family of states, the two-qubit example shows that in general 
the convex characteristic curve will give a lower bound for 
the mixed-state tangle but not necessarily a tight one.

\subsection{Mixed three-qubit states with GHZ and W components} 

For three qubits, the GHZ state and the W state 
are given by Eqs.~\eqref{ghz} and \eqref{wstate}, respectively.
We note that the 3-tangle of the GHZ state is maximal: $\tau_3(GHZ)=1$ while
it vanishes for the W state $\tau_3(W)=0$~\cite{Coffman00}. Moreover, we have
$\langle GHZ|W\rangle=0$.

In this example, we consider the family of mixed states 
\beq
    \rho(p)\ =\ p\ \pi_{\mathrm{GHZ}} \ +\ (1-p)\ \pi_{\mathrm W} \ \ ,
\label{mixedstate}
\eeq
where $p$ is a real number and $0\le p\le 1$. Our goal is to find the 
3-tangle $\tau_3(\rho(p))$ for all values of $p$.

According to our discussion
above, all elements of the optimal decompositions of $\rho(p)$
are linear combinations of $\ket{GHZ}$ and $\ket{W}$
\beq
   \ket{Z(p,\varphi)}\ =\ \sqrt{p}\ket{GHZ}\ -\ e^{i\varphi}\sqrt{1-p}\ket{W}
        \ \ .
\label{zstate}
\eeq
Let us first find the corners of the zero-polytope. 
By using the formula for the 3-tangle Eq.~(\ref{def-3tangle}) for
the states $\ket{Z(p,\varphi)}$ we obtain
\beq
     \tau_3(Z(p,\varphi))\ =
     \ \left|p^2-\frac{8\sqrt{6}}{9}\sqrt{p(1-p)^3}e^{3i\varphi}\right|\ =\ 0\ \ .
\label{3tangle_zstate}
\eeq
This equation is obeyed by the W state ($p=0$) as well as by the 
three states
\beq
   \ket{Z_0^{j}} = \sqrt{p_0}\ket{GHZ} \ -\ e^{\frac{2\pi ij}{3}} \sqrt{1-p_0}\ket{W}
\eeq
with $j=0, 1, 2 $ and
\beq
            p_0\ =\ \frac{4\sqrt[3]{2}}{3+4\sqrt[3]{2}} \ =\ 0.626851\ldots
\label{p0}
\eeq
These four states define the zero-polytope 
$S_0$ for the family $\rho(p)$ 
(which in this case is a simplex; see also Fig.~\ref{bloch}).
We note that all states inside $S_0$ have optimal decompositions of length four while the states
on the surface of $S_0$ have three-vector optimal decompositions. It is worthwhile mentioning 
that we have found many more
mixed states with vanishing 3-tangle than just those that belong 
to the family $\rho(p)$. Indeed, as explained above, $S_0$ contains all
density matrices in ${\cal R}_\rho$ with zero 3-tangle.

Interestingly, it is possible to determine the complete convex roof for $\rho(p)$ as
we will show now. To this end, we need to consider the corresponding convex characteristic 
curve. 
\begin{figure}[h]
\resizebox{.4\textwidth}{!}{\includegraphics{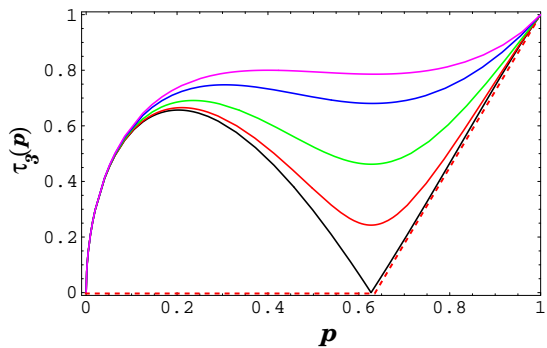}}

\caption{The 3-tangle of the states $\ket{Z(p,\varphi)}$ in 
         Eq.~(\ref{zstate}) for various values of $\varphi=\gamma  \cdot \frac{2\pi}{3}$ 
        (from top to bottom: $\gamma=1/2,1/3,1/5,1/10,0$).
      }
\label{GHZ-W}
\end{figure}
We have plotted $\tau_3(Z(p,\varphi))$
in Fig.~\ref{GHZ-W} and its minimum (solid black line for $\varphi=0$)
\beqa
\tilde{\tau}_{3}(p)\ &=&\ \min_{\varphi} \tau_3(Z(p,\varphi))
\\
                     &=&\ 
                \left|    p^2-\frac{8\sqrt{6}}{9}\sqrt{p(1-p)^3} \right| \ \ .
\label{tildetau3}
\eeqa
The function $\tilde{\tau}_3(p)$ is not  convex for $0\le p\le p_0$ and $0.825...\le p\le 1$.
Its function convex hull, i.e., the convex characteristic curve for $\rho(p)$ in 
Eq.~(\ref{mixedstate}) is given by
\beq
\tau^{\ast}_3(p) = \left\{ \begin{array}{ll}
                        0 \ \ , & 0\leq p\leq p_0     \\ 
        \tilde{\tau}_3(p)\ \ ,  & p_0\leq p \leq p_1  \\
     1 - (1-p)\left(\frac{3}{2}+\frac{1}{18} \sqrt{465} \right)
         \ , & p_1\leq p \leq 1
                     \end{array}
              \right.
\eeq
where 
\beq
           p_1\ =\ \frac{1}{2}\ +\ \frac{3}{310}\sqrt{465}\ =\ 0.70868\ldots
\label{eqp1}
\eeq

As we have explained in the preceding section $\tau^{\ast}(p)$
represents a lower bound for $\tau(\rho(p))$, and it coincides with the 3-tangle
for those values $p$ for which the corresponding $\tau^{\ast}(p)$ 
is realized by at least one decomposition. 
For the states $\rho(p)$ it is not difficult 
to specify a decomposition with an average 3-tangle equal to 
$\tau^{\ast}(p)$~(see also Ref.~\cite{LOSU}). 

In the interval $p_0\leq p \leq p_1$ we have the ``star-shaped'' 
three-vector decomposition  
 $\cal Z$
\beqa
   \rho(p)\ & = &\ \frac{1}{3}( 
             \ket{Z^{0}}\!\bra{Z^{0}}+
             \ket{Z^{1}}\!\bra{Z^{1}}+
             \ket{Z^{2}}\!\bra{Z^{2}})
   \label{decomp_p0}
   \\
   \ket{Z^{0}}\ & = &\ \sqrt{p}\ket{GHZ}\ -\ \sqrt{1-p}\ket{W}
   \nonumber\\
   \ket{Z^{1}}\ & = &\ \sqrt{p}\ket{GHZ}\ -\ 
                                   e^{\frac{2\pi i}{3}}\sqrt{1-p}\ket{W}
   \label{decomp_p0_vecs}\\
   \ket{Z^{2}}\ & = &\ \sqrt{p}\ket{GHZ}\ -\ 
                                   e^{\frac{4\pi i}{3}}\sqrt{1-p}\ket{W}
   \ \ .
   \nonumber
\eeqa

For $p\leq p_0$ a four-vector decomposition is optimal that combines
the decomposition $\cal Z$ in Eq.~(\ref{decomp_p0}) for $p_0$ with the W state:
\beq
 \rho(p)\ =\ \frac{p}{p_0}\ \rho(p_0)\ +\ 
                        \left(1-\frac{p}{p_0}\right)\ \pi_{\mathrm{W}}
   \ \ .
\label{decomp_zero}
\eeq
Finally, for $p_0 \leq p \leq 1$, we need to combine the decomposition $\cal Z$
for $p_1$ with the GHZ state:
\beq
 \rho(p)\ =\ \left(\frac{1-p}{1-p_1}\right)\ \rho(p_1)\ +\ 
                        \left(\frac{p-p_1}{1-p_1}\right)\ \pi_{\mathrm{GHZ}}
   \ \ .
\label{decomp_p1}
\eeq
We stress that, as the convex characteristic curve $\tau^{\ast}_3(p)$ 
gives a strict lower bound for $\tau_3(\rho(p))$, 
the existence of decompositions
attaining these values means that we have found the convex roof analytically.
It is worth noticing that the above given optimal decomposition
is unique.

\begin{figure}[h]
\resizebox{.4\textwidth}{!}{\includegraphics{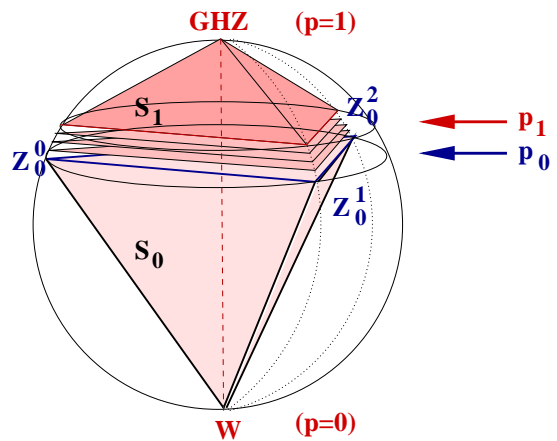}}

\caption{Bloch sphere for the two-dimensional space spanned
         by the GHZ state and the W state.
         The simplex $S_0$ contains all states with vanishing 3-tangle.
         The ``leaves'' between $p_0$ and $p_1$ represent sets of
         constant 3-tangle.
         In the simplex $S_1$, the 3-tangle is an affine function
         (for further explanation, see text).
      }
\label{bloch}
\end{figure}
The extension of the analytic convex roof to the entire simplices
spanned by the decomposition vectors (see Fig~\ref{bloch}) follows from the
theorem that the vectors in an optimal decomposition are also optimal for the 
entire simplex they span (this applies also to states forming a subset of 
an optimal decomposition)~\cite{Uhlmann98}.

\subsection{Generic example for a rank-2 mixed two-qubit state} 

In the previous example we have seen that studying the convex characteristic curve 
may lead to a complete solution for the convex roof of a family of mixed states.
In general, however, this may not be expected. 
The reason why the characteristic curve will typically fail
to give the exact convex roof
is because the phases $\varphi_{j}$ for which the minimum
in Eqs.~\eqref{tautilde}, \eqref{char-curve-gen} is reached, do not necessarily 
admit a decomposition of $\rho$.

Nevertheless one obtains 
valuable information about a given mixed state. In order to illustrate this,
we describe a rather generic two-qubit case. Let us consider the states
\begin{eqnarray}
       \ket{I}\ &=& \ \frac{1}{\sqrt{5}}\ (2\ket{00}\ +\ \ket{11})
\label{stateI}
\\
  \ket{II}\ &=&\ \frac{1}{\sqrt{6}}\ ( \ket{00} + \ket{01}  - 2\ket{11} )
\label{stateII}
\end{eqnarray}
and their convex combinations
\beq
    \rho_2(p)\ =\ p\ \pi_{I} \ +\ (1-p)\ \pi_{II} \ \ ,
\label{mixedstate2bit}
\eeq
We note that $\langle I|II\rangle=0$. 
This example can easily be treated analytically by 
using Wootters' method~\cite{Wootters98}. 

\begin{figure}[ht]
\includegraphics[width=.4\textwidth]{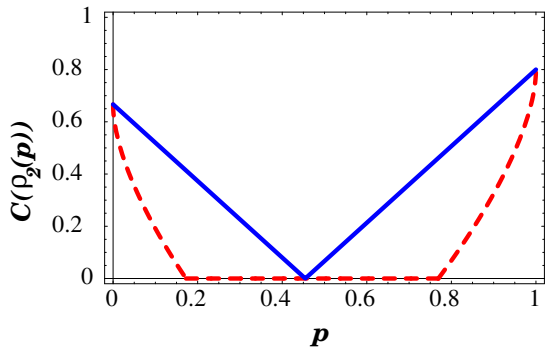}
\caption{Solid (blue) line: exact solution for the concurrence $C(\rho_2(p))$ according
         to Eqs.~(\ref{stateI})--(\ref{mixedstate2bit}). For comparison: the 
         convex characteristic curve of $\rho_2(p)$ (dashed red line; cf.\ also
         Fig.~5). The exact concurrence $C(\rho(p))$ has a zero for \mbox{$\tilde{p}_0=5/11$.}}
\label{conc2}
\end{figure}

In particular, we find vanishing concurrence for $\tilde{p}_0=5/11$ 
(see Fig.~\ref{conc2}).

Now let us first evaluate the zero-polytope  (which is a line in the Bloch sphere in
this case). The equation 
\beq
         C(\sqrt{p}\ket{I}+e^{i\varphi}\sqrt{1-p}\ket{II})=0
\label{conczero}
\eeq
has two real solutions for $p$: $\tilde{p}_1=5/29$ with $\varphi=\pi$ 
and $\tilde{p}_2=10/13$ with $\varphi=0$. 
>From this we can conclude immediately that there is exactly one
value of $p$ for which the mixed-state concurrence of $\rho_2(p)$ vanishes: it
is the intersection point of the zero-polytope  with the line in the Bloch
sphere that represents the family $\rho_2(p)$. Straightforward algebra
shows that this intersection occurs at \mbox{$\tilde{p}_0=5/11$}.

We see that the zero-polytope indeed gives us exact information 
about a subset of the family $\rho_2(p)$ only
(here: about a single element).
Note however that this subset can even be empty in the
most general case, when e.g. the wave function coefficients of
\eqref{stateI} have a non-zero relative phase modulo $\pi$.

\begin{figure}[h]
\vspace*{7mm}
\includegraphics[width=.4\textwidth]{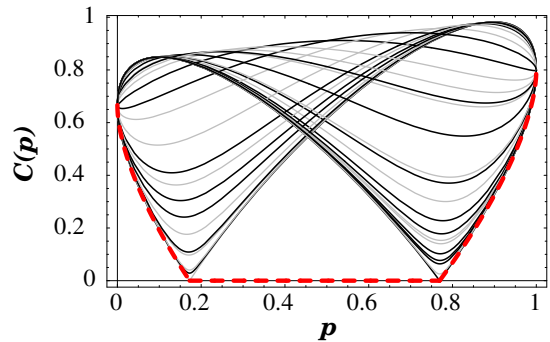}
\caption{Concurrence for superpositions of states $\ket{I}$ and $\ket{II}$ (solid gray
         lines) and convex characteristic curve $C^{\ast}(p)$ 
         (dashed red line). The convex characteristic curve
         coincides with the abscissa (i.e., $C^{\ast}(p)\equiv0$)
         between the two zeros of Eq.~(\ref{conczero}), $\tilde{p}_1=5/29$
         and $\tilde{p}_2=10/13$.}
\label{kenn-2}
\end{figure}
In Fig.~5 we show how the convex characteristic curve of $\rho_2(p)$
arises from the concurrence of superpositions of the states $\ket{I}$ and
$\ket{II}$. 
Analytically, the convex characteristic curve reads:
\beq
        C^{\ast}(p) = \left\{ \begin{array}{ll}
        C_-(p)\ \ ,  & 0\leq p \leq \tilde{p}_1  \\
                        0 \ \ , & \tilde{p}_1\leq p\leq \tilde{p}_2     \\ 
        C_+(p)\ \ ,  & \tilde{p}_1\leq p \leq 1  \\
                     \end{array}
              \right.
\eeq
where $C_{\pm}(p)=C(\sqrt{p}\ket{I}\pm\sqrt{1-p}\ket{II})$. 
Clearly, $C^{\ast}(p)$ must vanish for all values of $p$ between
the  zeros according to Eq.~(\ref{conczero}). We observe (cf.~Fig.~4) that
$C^{\ast}(p)$ is indeed a lower bound to the exact concurrence. However,
the latter is strongly underestimated by the convex characteristic curve
for almost all $p$ values.

\section{Discussion}
We have introduced two concepts that serve to facilitate the investigation of
the convex roof of tangles for multipartite mixed qubit states. Both concepts
are based on the tangle of pure-state superpositions, that is, of those pure
states that lie in the range of the original density matrix.

The zero-polytope is formed by the convex set of density 
matrices with vanishing tangle
in the range of the original mixed state under consideration. 
It can always be determined exactly. Then, the remaining task is the 
calculation
of the convex roof for all remaining density matrices with non-zero tangle.

To this end, we have proposed the concept of the convex characteristic 
curve (or manifold).  It leads to a non-trivial and possibly tight 
lower bound for the tangle of a given density matrix. 
In particular, it coincides with the convex roof if there is a decomposition
of the original density matrix into pure states whose tangles lie on the 
convex characteristic curve.

In order to illustrate the power of these concepts we have discussed two simple examples --
three-qubit mixed states with GHZ and W components~\cite{LOSU}, 
and a generic rank-2 two-qubit state.
For the GHZ/W mixtures the convex characteristic curve shows analytically
{\em i)} that the results of Ref.~\cite{LOSU} indeed represent 
the convex roof for that
family of states, and {\em ii)} provide an example where the convex 
characteristic curve
provides the complete solution for a convex roof -- 
namely, due to the fact that for each
point of the convex characteristic curve there exists a 
decomposition that attains 
this lower bound. On the other hand, the two-qubit example 
displays that in a more general 
case, one may obtain non-trivial estimates for the convex roof. 
However, as in that case
there hardly exists an optimal decomposition 
with all its elements on the convex 
characteristic curve, on must not expect that the obtained 
estimates for the convex
roof are close to the exact value.

While, in principle, both concepts can be generalized to mixed 
states of arbitrary 
rank, their application is particularly adapted to the case of rank-2 states.
\vspace*{5mm}
\section{Acknowledgments}

We would like to thank C.\ Eltschka and R.\ Lohmayer
for stimulating discussions and helpful comments. 
This work was supported by the EU 
RTN grant HPRN-CT-2000-00144 
and the Sonderforschungsbereich 631 
of the German Research Foundation. J.S. receives support from
the Heisenberg Programme
of the German Research Foundation.

\end{document}